\documentclass[%
 aip,
 jmp,%
 amsmath,amssymb,
 reprint,%
]{revtex4-1}

\usepackage{graphicx}
\usepackage{dcolumn}
\usepackage{bm}

\begin{document}
\preprint{AIP/123-QED}

\title[Transition between half-metal and ferromagnetic semiconductor induced by silicon vacancy in epitaxial silicene]{Transition between half-metal and ferromagnetic semiconductor induced by silicon vacancy in epitaxial silicene}

\author{Yan Qian}
 \email{qianyan@njust.edu.cn}
\author{Erjun Kan}
\author{Kaiming Deng}
\author{Haiping Wu}%
 \email{mrhpwu@njust.edu.cn}

\affiliation{
Department of Applied Physics, Nanjing University of Science and
Technology, Nanjing 210094, China
}%

\date{\today}

\begin{abstract}
Since the inevitability in experimental synthesis, defects show great importance to many materials. They will deeply regulate the properties of the materials, and then affect the further applications. Thus, exploring the effects of defects on the properties of materials is desired. Here, by using first-principles calculations, we systematically studied the effect of silicon vacancy defects on the properties of silicene generated on N-terminated cubic boron nitride (111) surface. It is found that the introduction of silicon vacancy would trigger transition between half-metal and ferromagnetic semiconductor. With small vacancy ratios of 1/36 and 1/24, the ground-state of the samples would behave as ferromagnetic semiconductors, and the band gaps are about 1.25 and 0.95 eV, respectively. When the vacancy ratio is increased up to 1/6, the sample would turn into a ferromagnetic half-metal with a half-metallic gap of around 0.15 eV. The change of the electronic structure of the samples is driven by the different electron transfer between silicon layer and substrate, i.e., there will be different amount of electrons transferred from the silicon layer to the substrate when the vacancy ratio is altered. This work would open a new way to regulate the properties of materials and extend applications in nanoelectronic field.

\end{abstract}

\maketitle
\section{Introduction}
With the development of electronic industry, increasing the integration level is still the goal researchers pursuing. Low-dimensional materials can satisfy this demand at least on scale.\cite{Boon,Tang,Rao} Thus, a great deal of work is focused on designing different kinds of low-dimensional materials, such as elemental two-dimensional(2D) nanosheets,\cite{AJ} 2D atomically thin transition metal dichalcogenides (TMDs),\cite{SM} 2D monochalcogenides,\cite{KX,DA} and so on. Among these 2D materials, low-dimensional Si-based materials show specially importance to the modern electronic fields, since there is an adequate amount of silicon on the earth and silicon-based materials are still the main constituents in the most electronic components nowadays. For this reason, a number of 2D silicon allotropes beyond silicene\cite{CS1} habe been reported, for example, layered silicite,\cite{CS2} silicoctene,\cite{HP} multilayer silicene,\cite{RY,LJ} etc. However, only silicene and bilayer silicene were experimentally synthesized up to our knowledge.\cite{BL,LC,BF,PV,AF,LM,LH,AS,TA,RY}

Moreover, to manufacture an electronic component, the low-dimensional materials must adhere onto some semiconducting or insulating substrates. Unfortunately, almost all the silicenes are experimentally grown on metallic substrates.\cite{BL,LC,BF,PV,AF,LM,LH,AS,TA} Therefor, for the practical applications, silicene must be transferred from the metallic substrate to the nonmetallic destination, and this process would induce many defects in samples, such as wrinkles, damages, contaminations, etc. Furthermore, the destruction of the material would greatly change its intrinsic properties. Additionally, the interaction between silicene and the different substrates is another factor possibly modifying the properties of silicene. On the contrary, this interaction also can be used as an effective means to tune the properties of the adhered 2D materials. Thus, searching for a suitably nonmetallic substrate has been a promising way for property engineering, and generating silicene directly on nonmetallic substrates has attracted considerable attention. A few works have been reported such as the following. Several 2D silicon allotropes were generated on cubic boron nitride (c-BN) substrate in our previous work,\cite{HP1,HP2} silicene was grown on Al$_{2}$O$_{3}$ substrate and exhibited semiconducting nature with a band gap of around 0.40 eV,\cite{Chen} and so on.
However, few work was focused on the effect of defect on the properties of 2D silicon materials generated on nonmetallic substrates, although the defect is inevitable in experimental synthesis. Besides, introducing defect also can be used as a technology to tune the properties of the materials. Thus, by using first-principles calculations in this work, we introduced several types of silicon vacancy defects into silicene generated on the N-terminated c-BN (111) surface, and some interesting properties were found.

\section{Computational methods}
During the calculations, the original structure is taken from the previous work,\cite{HP2} and then three types of silicon vacancy defect (i.e. three different vacancy ratios) are introduced, as plotted in Fig. 1. To design the samples with different vacancy ratios, three different supercells (2x3, 2x2, and 1x1 cell, respectively) are employed and all planted with one silicon vacancy. Thus, the ratios of vacancy site to silicene lattice sites are 1/36, 1/24, and 1/6, respectively, and the three corresponding samples are named as Sample$_{1/36}$, Sample$_{1/24}$, and Sample$_{1/6}$. The structure relaxations and electronic structure calculations are carried out in the framework of density functional theory (DFT) within generalized-gradient approximations employing the PerdewBurke-Ernzerhof (PBE)\cite{Perdew} exchange correlation functional and projector-augmented-wave (PAW)\cite{Kresse1} potentials as performed in VASP.\cite{Kresse2} To ensure enough accuracy, the k-point density and the plane waves cutoff energy are increased until the change in the total energy is less than 10$^{-5}$ eV, and the Brillouin-zone (BZ) integration is carried out using 5$\times$3$\times$1, 5$\times$5$\times$1, 11$\times$11$\times$1 Monkhorstack grid in the first BZ for Sample$_{1/36}$, Sample$_{1/24}$, and Sample$_{1/6}$, respectively, and the plane waves of kinetic energy is up to 500 eV. Structure relaxations are carried out until the Hellmann-Feynman force on each atom is reduced by less than 0.01 eV{\AA}$^{-1}$.

\section{Results and discussions}
Before any further investigations on
\begin{figure}[htbp]
\centering
\includegraphics[width=8.5cm]{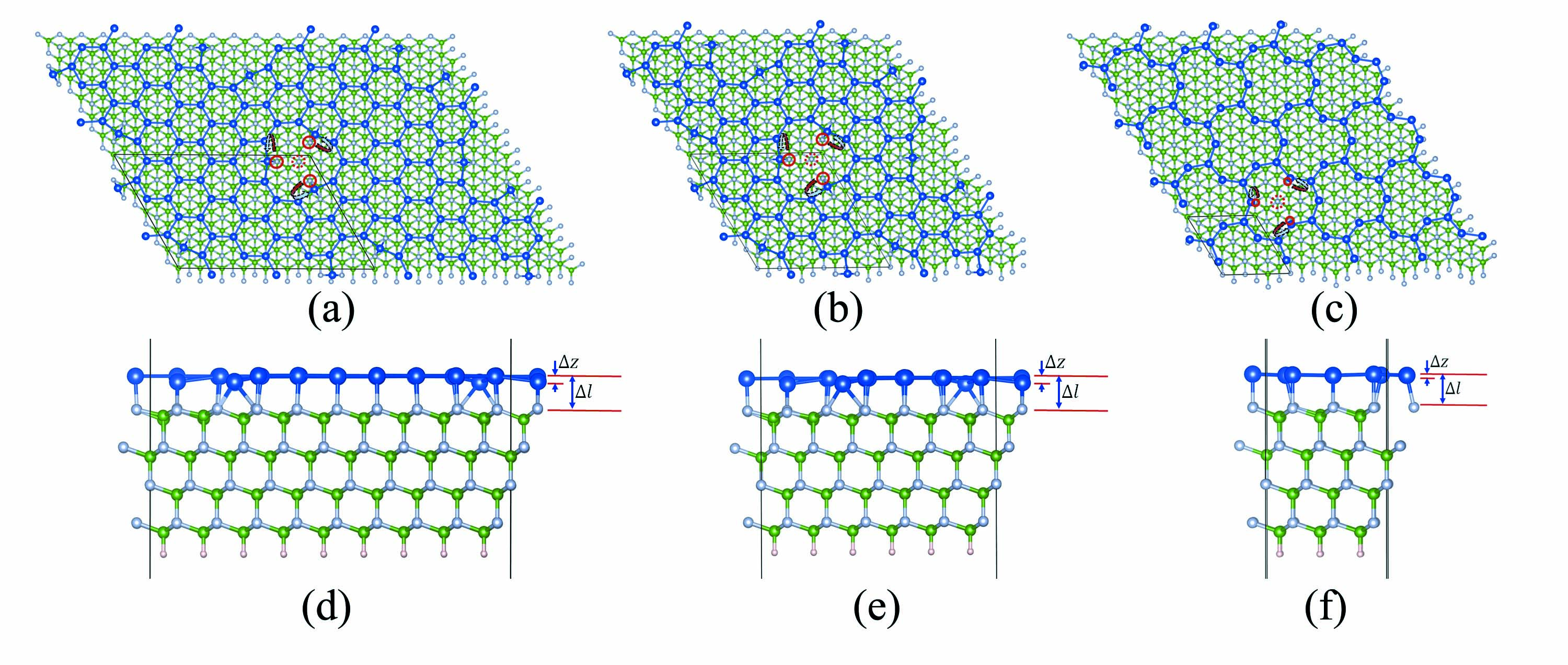}
\caption{(Color online). The optimized structure of Sample$_{1/36}$, Sample$_{1/24}$, and Sample$_{1/6}$. (a), (b) and (c) are the top views, and (d), (e), and (f) are the side views. The cells in solid line quadrangle are taken as the supercells for calculations. The blue, green, and gray balls are Si, B, and N atoms, respectively, and the brown ball presents H atom. The solid red circles present the original positions of Si atoms nearest-neighboring the vacancy sites named as Si$_{nv}$, and the dot red circles are the positions of silicon vacancies. The other Si atoms are named as Si$_{ot}$.}
\label{fig:Figure1}
\end{figure}
the three samples, the magnetic phases of their ground-state structures were explored firstly. The calculated total energies of both phases are listed in TABLE I. It shows that the total energies of the ferromagnetic phase have about 0.13, 0.21, and 0.07 eV lower than those of the nonmagnetic one for Sample$_{1/36}$, Sample$_{1/24}$, and Sample$_{1/6}$, respectively. This implies that the ferromagnetic ground-state structure is independent on the silicon vacancy ratio.
\begin{table}[!hbp]
\caption{Vacrat presents the silicon vacancy ratio. TE$_{FM}$ and TE$_{NM}$ are the total energies for three samples with ferromagnetic and nonmagnetic phases, respectively. Here, the total energies of the ferromagnetism are shifted to zero. $\Delta$E is for the energy difference per cell between the two magnetic phases. Mag$_{Nbv}$, Mag$_{Not}$, Mag$_{Sinv}$, and Mag$_{Siot}$ are for the magnetic moments located on N$_{bv}$, N$_{ot}$, Si$_{nv}$, and Si$_{ot}$ sites, respectively.}
\begin{tabular}{p{0.9cm}p{0.85cm}p{0.9cm}p{0.60cm}p{1.05cm}p{1.05cm}p{1.10cm}p{1.0cm}}
\hline
\hline
Vacrat  &TE$_{FM}$ & TE$_{NM}$ &  $\Delta$E & Mag$_{Nbv}$ & Mag$_{Not}$ & Mag$_{Sinv}$ &  Mag$_{Siot}$  \\
\hline
1/36     & 0   & 0.13 & 0.02 & 0.43 &  0.11 & 0.00  & 0.00\\
\hline
1/24     & 0   & 0.21 & 0.05 & 0.42 &  0.00 & 0.00  & 0.00 \\
\hline
1/6      & 0   & 0.07 & 0.07 & -0.04&  0.00 & 0.07  & 0.13 \\
\hline
\hline
\end{tabular}
\end{table}
Associated with the ferromagnetism for the sample without vacancy defect,\cite{HP2} it can be concluded that silicene generated on N-terminated c-BN (111) surface would always retain ferromagnetic phase, regardless of the introduced vacancy. In the following, the properties of the ferromagnetic ground-state phases for the samples are discussed if not specially noted. Besides, the table also indicates that, with the increasing of vacancy ratio, the total energy per cell of the ferromagnetic phase relative to the nonmagnetic one increases from 0.02 up to 0.07 eV, this trend denotes that the Curie temperature of the sample would increase with increasing the silicon vacancy ratio.

In order to investigate the magnetic property in detail, the electronic charge density difference between the spin-up and spin-down channels is plotted in Fig. 2.
\begin{figure}[htbp]
\centering
\includegraphics[width=8.5cm]{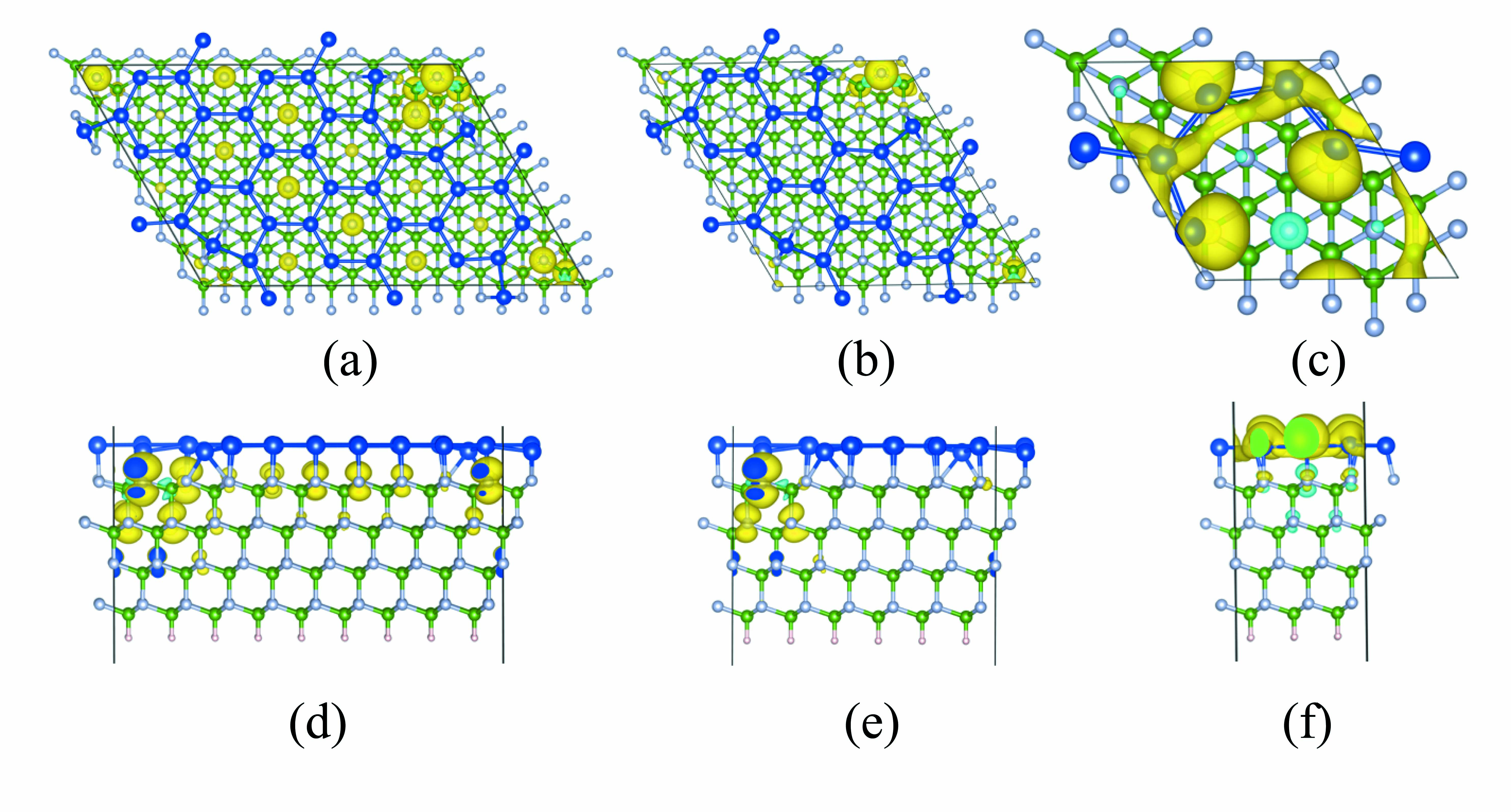}
\caption{(Color online). The electronic charge density difference between the spin-up and spin-down channels of Sample$_{1/36}$, Sample$_{1/24}$, and Sample$_{1/6}$, respectively. (a), (b) and (c) are the top views, and (d), (e), and (f) are the side views.}
\label{fig:Figure2}
\end{figure}
For Sample$_{1/36}$, the major magnetic moments of $\sim$0.43 $\mu$$_{B}$ are located on N atoms right below the vacancy sites in the surface BN layer (named as N$_{bv}$), and some small moments of $\sim$0.11 $\mu$$_{B}$ are situated on the other N atoms not bonded with Si atoms in the same BN layer (named as N$_{ot}$). The moments on N$_{ot}$ sites are a little smaller than that of $\sim$0.12 $\mu$$_{B}$ on the corresponding N sites in the sample without defect.\cite{HP2} When the vacancy ratio is 1/24, the major magnetic moments are of $\sim$0.42 $\mu$$_{B}$ located on N$_{bv}$ atoms, but the magnetic moments are almost disappeared on the N$_{ot}$ sites. As long as the vacancy ratio is increased up to 1/6, the major magnetic moments are attributed to Si atoms, and the very small moments are situated on N$_{bv}$ atoms antiparallel to those of Si atoms. In detail, the values are $\sim$ 0.13 and 0.07 $\mu$$_{B}$ on the Si$_{ot}$ and Si$_{nv}$ atoms, respectively, and it is $\sim$ 0.04 $\mu$$_{B}$ on N$_{bv}$ atoms. From the above data, it obviously shows that the magnetic moment on Si sites would increase with the increase of vacancy ratio, but it is oppsite for N atoms. This magnetic moment trend can be clearly explained by the electron transfer from Si to N atoms. The introduction of silicon vacancy in the sample would lead to the appearance of dangling bonds in Si$_{nv}$ and N$_{bv}$ atoms, besides those already existed in N$_{ot}$ atoms in the sample without defect. This indicates that there would be unpaired electrons in Si$_{nv}$ and N$_{bv}$ atoms. Associated with the stronger electronegativity of N atom, some electrons would transfer from Si$_{nv}$ to N atoms. The more silicon vacancy introduced, the more electrons would transfer from Si to N atoms. As long as the dangling bonds in N atoms are saturated, the remained unpaired electrons would distribute mainly in Si atoms if increasing the vacancy ratio on end, which causing the magnetism of Si atoms.

\begin{table}[!hbp]
\caption{$\Delta$l is for the diastances (in {\AA}) between c-BN (111) surface and silicon layer, $\Delta$z is for the buckling of Si layer along $<$001$>$ direction, and $\Delta$r is for the displacement of Si$_{nv}$. l$_{1}$$_{(Si-Si)}$ is for the lengths of Si$_{ot}$-Si$_{nv}$ bonds, and the lengths of Si$_{ot}$-Si$_{ot}$ bonds are expressed by l$_{2}$$_{(Si-Si)}$. }
\begin{tabular}{p{1.1cm}p{1.2cm}p{1.2cm}p{1.0cm}p{1.3cm}p{1.6cm}}
\hline
\hline
Vacrat &$\Delta$l & $\Delta$z & $\Delta$r  & l$_{1}$$_{(Si-Si)}$  & l$_{2}$$_{(Si-Si)}$ \\
\hline
1/36 & 1.85 &  0.33 &  1.10 &  2.42 & 2.53 \\
\hline
1/24 & 1.83 &  0.33 & 1.08  & 2.41 & 2.52 \\
\hline
1/6 &  1.82 &  0.06 &  0.44 &  2.36 & $\diagup$ \\
\hline
\hline
\end{tabular}
\end{table}

 Due to the strong interaction between the Si and N atoms, the generated silicene exhibits absolutely flat structure when no defect is introduced, and all the Si atoms are stationed right above the N atoms. While this structural character would be changed as long as the silicon vacancy is imported, and some structural data of these three samples are listed in TABLE II. The change happens mainly on the Si$_{nv}$ sites, and the Si$_{ot}$ atoms are still situated right above the N atoms. The Si$_{nv}$ atoms would move away from the original position in the radial direction if we take the vacancy site as the center of a circle. For Sample$_{1/36}$ and Sample$_{1/24}$, the displacements of Si$_{nv}$ atoms are $\sim$ 1.10 and 1.08 {\AA} in (001) plane, respectively. While the Si$_{nv}$ atoms shift only $\sim$ 0.44 {\AA} for Sample$_{1/6}$, this value is much smaller than those of the other two samples. It denotes that the displacement of Si$_{nv}$ in (001) plane will decrease with the increasing of the vacancy ratio, and this displacement trend can explained by the transferring of electron, which is in accordance with the interpretation of magnetic moment tendency. With increasing of the vacancy ratio, there will be more surface N atoms not bonded with Si atoms, leading to more electrons transferred from Si$_{nv}$ to surface N atoms. On the other hand, Si$_{nv}$ atoms are bonded with two nearest-neighboring N atoms (this is confirmed by the electron localization functions in the following). With more electrons transferred to N atoms, the dangling bond of the next-nearest-neighboring N$_{ot}$ atom would be suppressed gradually, so that the Si$_{nv}$-N$_{ot}$ bond would weaken and vanish finally. The weakening of Si$_{nv}$-N$_{ot}$ bond leads to the shorter displacement of Si$_{nv}$ in (001) plane. In the same sample, the different interaction between Si and different surface N atoms would bring about the buckling characteristics of silicon layer along $<$001$>$ direction, instead of the absolutely plane structure in generated silicene without defect.\cite{HP2} The surface N atoms interact more strongly with Si$_{nv}$ than Si$_{ot}$ atoms, since the Si$_{nv}$ atoms have much more electrons to bond with N atoms than Si$_{ot}$ atoms. The buckling also shows decreasing tendency from 0.33 to 0.06 {\AA} with increasing the vacancy ratio, caused by the weakening interaction between Si$_{nv}$ and two nearest-neighboring N atoms. The distances between the silicon monolayer and the substrate surface of the three samples are all around 1.83 {\AA}, similar with $\sim$1.83 {\AA} of the sample without defect.\cite{HP2} For the structure of silicon layer, the lengths of Si-Si bonds of the six-membered rings are remained of $\sim$ 2.53 {\AA}. While the lengths of the other Si-Si bonds including Si$_{nv}$ atoms are slightly shorter and become from 2.42 to 2.36 {\AA}, it illustrates that this Si-Si bond length would decrease with increasing the vacancy ratio. This trend originates from the change of Si$_{nv}$-N interaction with different silicon vacancy ratio.

To explore the interaction between Si and surface N atoms, the electron localization functions (ELF) is plotted in Fig. 3. From the top views of the three samples, all the Si-Si bonds behave as strong $\sigma$ state. Additionally, the Si-Si$_{nv}$ bonds simultaneously show the nature of $\pi$ state. All the side views show that there exist strong $\sigma$ bonds between Si and N atoms. Figs. 3(d) and 3(e) also show that Si$_{nv}$ atoms are bonded with two nearest-neighboring N atoms in Sample$_{1/36}$ and Sample$_{1/24}$ except in Sample$_{1/6}$. Additionally, the formation of strong bonds also interprets that the samples could remain stably independent of the ratio of silicon vacancy.
\begin{figure}[htbp]
\centering
\includegraphics[width=8.5cm]{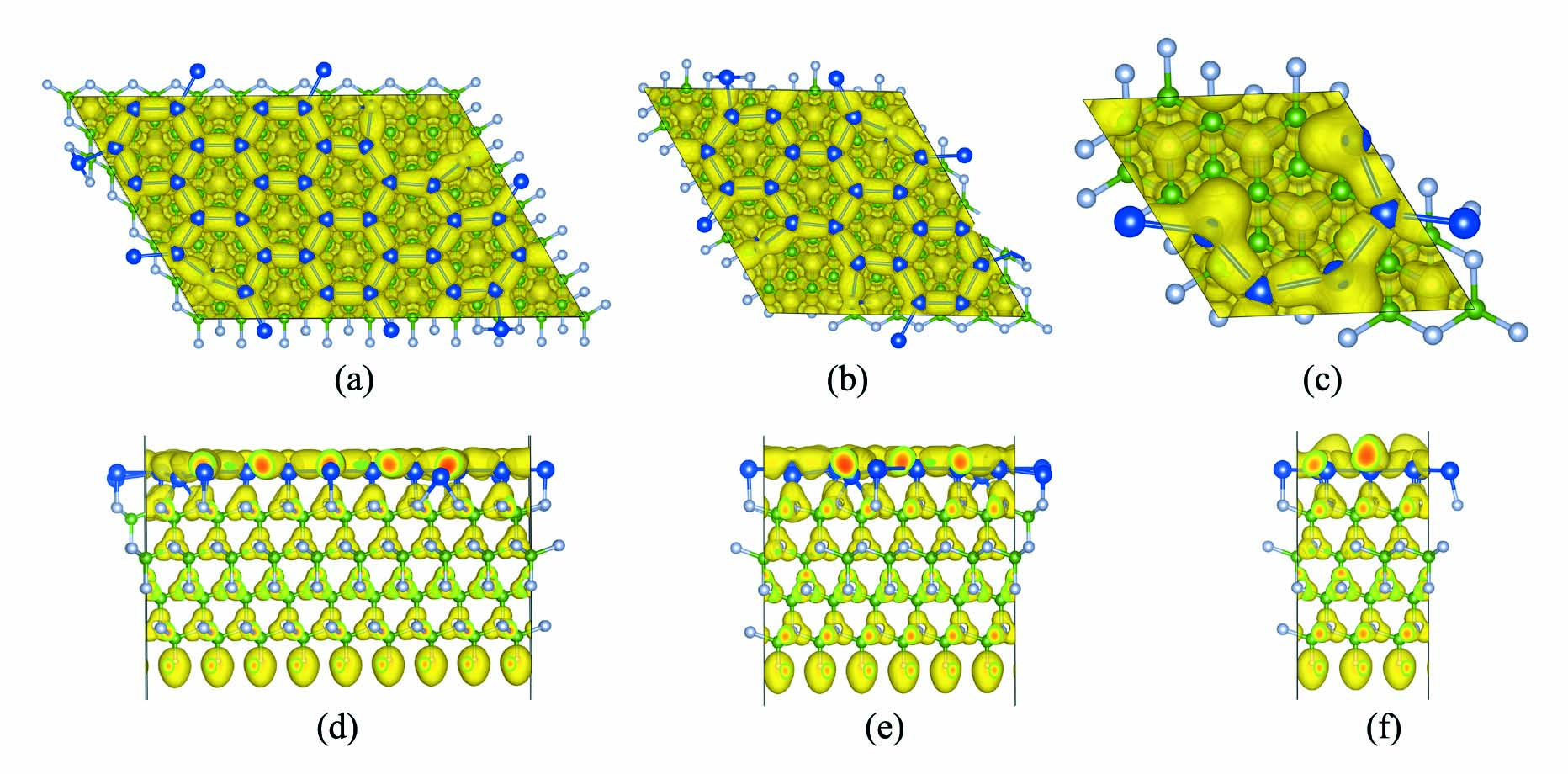}
\caption{(Color online). The electron localization functions of Sample$_{1/36}$, Sample$_{1/24}$, and Sample$_{1/6}$, respectively. (a), (b) and (c) are the top views, and (d), (e), and (f) are the side views.}
\label{fig:Figure3}
\end{figure}

The electronic properties are investigated finally. Figure 4(a) draws the total density of electronic states (DOS), and it demonstrates that Sample$_{1/36}$ and Sample$_{1/24}$ both exhibit semiconducting property, in that there are no electronic states located at the Fermi energy level (\textit{E}$_{F}$). The band gaps are $\sim$ 1.25 and 0.95 eV, respectively. The states located at 0.49 eV of Sample$_{1/24}$ can be regarded as a deep impurity level. This reveals that the ground-states of Sample$_{1/36}$ and Sample$_{1/24}$ are both ferromagnetic semiconductors. For Sample$_{1/6}$, the figure clearly displays that there are many electronic states situated at \textit{E}$_{F}$ in the spin-up channel, but no electronic states are located at \textit{E}$_{F}$ in the spin-down channel. This illustrates that Sample$_{1/6}$ is a half-metal, and the half-metallic gap (the minimal energy gap for a spin excitation) is $\sim$ 0.15 eV, this value is some larger than 0.11 eV of the sample without defect.\cite{HP2}  The partial density of electronic states (PDOS) near \textit{E}$_{F}$ are pictured in Figs. 4(b), 4(c), and 4(d). From Figs. 4(b) and 4(c), it shows that there are very few electronic states of N atoms situated above \textit{E}$_{F}$, indicating that few N 2\textit{p} orbitals are not completely occupied. The figures also show that there are some Si 3\textit{p} states located above \textit{E}$_{F}$.  Fig. 4(d) shows that all the N 2\textit{p} orbitals are below \textit{E}$_{F}$, and many Si 3\textit{p} states are situated at and above \textit{E}$_{F}$. This characteristics is clearly different with that of the sample without defect, although they both exhibit half-metallic property. For the sample without defect, the electronic states at \textit{E}$_{F}$ is mainly attributed to N 2\textit{p} orbitals. This trend of PDOS coordinates with the rule of electron transfer explained above. With increasing the silicon vacancy ratio, the dangling bonds of N atoms would be saturated due to the electrons transferred from Si atoms. On the other hand, the unpaired electrons of Si atoms would increase. Therefor, the Si 3\textit{p} states would substitute for N 2\textit{p} states at \textit{E}$_{F}$ as long as enough silicon vacancy is introduced.
\begin{figure}[htbp]
\centering
\includegraphics[width=8.5cm]{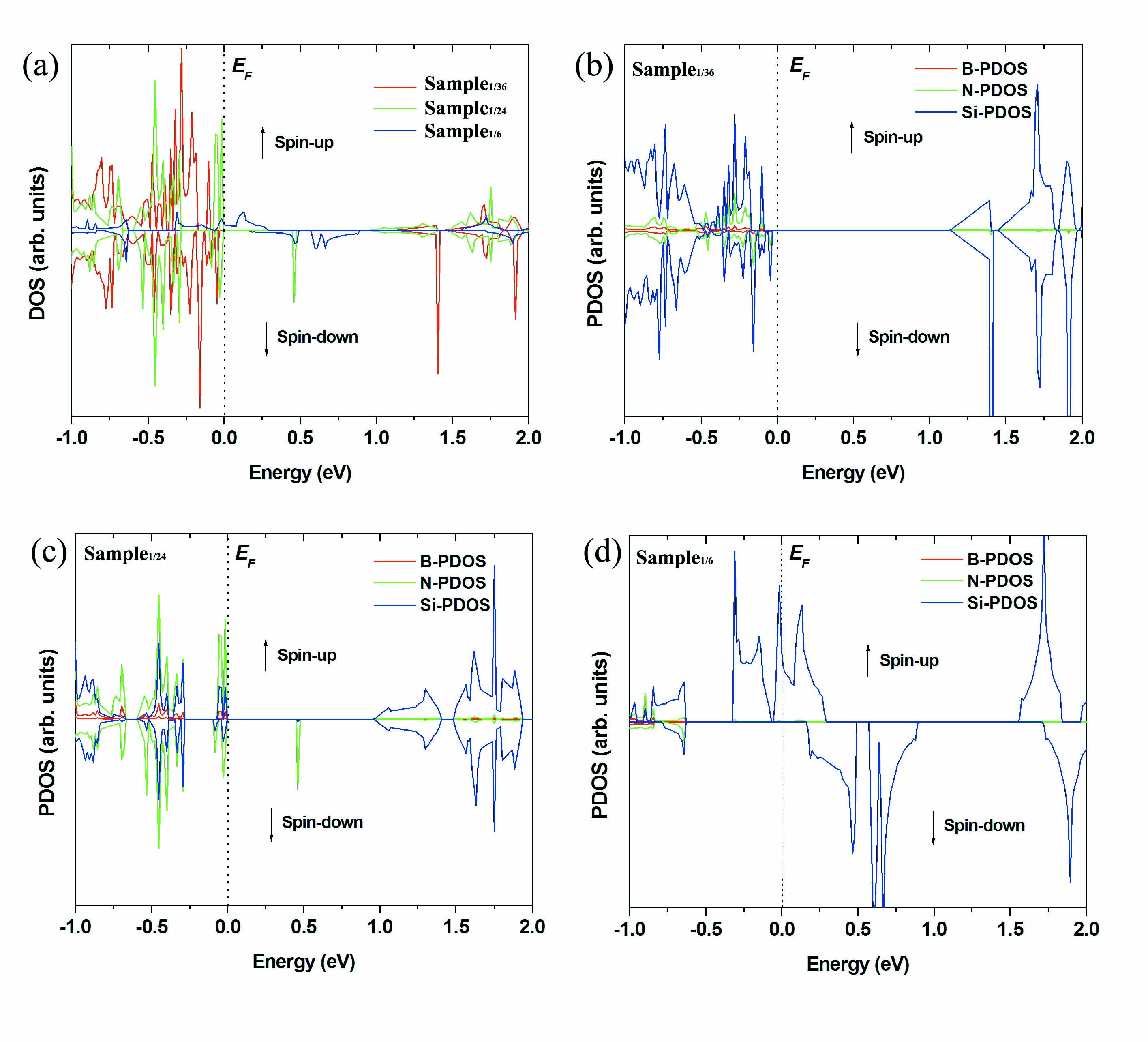}
\caption{(Color online). (a) The DOS of Sample$_{1/36}$, Sample$_{1/24}$, and Sample$_{1/6}$, respectively. (b), (c), and (d) are the PDOS for Sample$_{1/36}$, Sample$_{1/24}$, and Sample$_{1/6}$, respectively.}
\label{fig:Figure4}
\end{figure}

\section{Conclusions}
In summary, in order to regulate the property of silicene and extend its applications in modern electronic field, we introduced silicon vacancy in silicene generated on N-terminated c-BN (111) surface. As a result, the diverse electronic properties are tuned, such as ferromagnetic semiconductor and half-metal.
The electronic structure is dependent on the silicon vacancy ratio, so that we can obtain the needed electronic property for the corresponding applications by controlling the silicon vacancy ratio.

\vspace{1ex}
Acknowledgments

This work was supported by the National Natural Science Foundation of China (Grant no. 11404168).

Data Availability

The data that support the findings of this study are available from the corresponding author upon reasonable request.

\end{document}